**NOMAD: The FAIR Concept for Big-Data-Driven Materials Science**

Claudia Draxl[1,2] (*) and Matthias Scheffler[2,1,3]

[1] Physics Department, Humboldt-Universität zu Berlin, Germany

[2] Fritz-Haber-Institut der Max-Planck-Gesellschaft Berlin, Germany

[3] University of California Santa Barbara, Department of Chemistry and Biochemistry, Santa Barbara, USA

Data is a crucial raw material of this century, and the amount of data that has been created in materials science in recent years and is being created every new day is immense. Without a proper infrastructure that allows for collecting and sharing data (including the original data), the envisioned success of materials science and, in particular, Big-Data driven materials science will be hampered. For the field of computational materials science, the NOMAD (Novel Materials Discovery) Center of Excellence (CoE) has changed the scientific culture towards a comprehensive and FAIR data sharing, opening new avenues for mining Big-Data of materials science. Novel data-analytics concepts and tools turn data into knowledge and help the prediction of new materials or the identification of new properties of already known materials.

**Introduction**

The discovery of improved or even novel - not just new - materials or hitherto-unknown properties of known materials to meet a specific scientific or industrial requirement is one of the most exciting and economically important applications of high-performance computing (HPC) to date. The convergence of theoretical physics and chemistry, materials science and engineering, and computer science into computational materials science enables the modeling of materials, both existing materials and those that can be created in the future, at the electronic and atomic level. This also allows the accurate prediction of how these materials will behave at the microscopic and macroscopic levels, and so to understand their suitability for specific research and commercial applications. Computational high-throughput screening initiatives (for example, Refs. 1,2,3,4,5,6), significantly boosted by the Materials Genome Initiative[7,8] of the White House, meet this issue by computing the properties of many thousands of structures. Having a closer look, however, one realizes that such studies are inefficiently exploited, so far, as only a tiny amount of the information that is contained in all the computed data is being used. In fact, this drawback applies to high-throughput screening as well as to any individual theoretical and experimental investigation. Unfortunately, besides a few numbers, tables or graphs that appear in the resulting publication, the wealth of other information contained in the full research work is typically disregarded or even deleted.

Changing this situation and fully realizing a comprehensive *Data Sharing* is slow and more characterized by lip services from science politics and funding agencies than by commitments and support.[9] This slows the possible progress of scientific advancements. In this context and in the area of computational materials science, the NOMAD (Novel Materials Discovery Center of Excellence (CoE)[10] has assumed a pioneering role, considering all aspects of what is now called FAIR[11,12] handling of data: Data are **F**indable for anyone interested; they are stored in a way that they are easily **A**ccessible; their representation follows accepted standards[13], and all specifications are open – hence data are **I**nteroperable[14]. All this enables that data can be used for research questions that can be different from the purpose they have been created for; hence data are **R**e-purposable[15]. We illustrate the latter with an example: Let's assume a research team has investigated $TiO_2$ in view of heterogeneous catalysis where $TiO_2$ is an important support material. The results published in a research journal are – as not comprehensive – typically not useful for researchers who are interested in the same material but in a different context. Having all data in a normalized and code-independent form, as they are available in the NOMAD Archive[16], allows one to use it in a very versatile form. This data can be as useful for photovoltaics in inorganic/organic hybrid materials as for research on the color of $TiO_2$.[17]

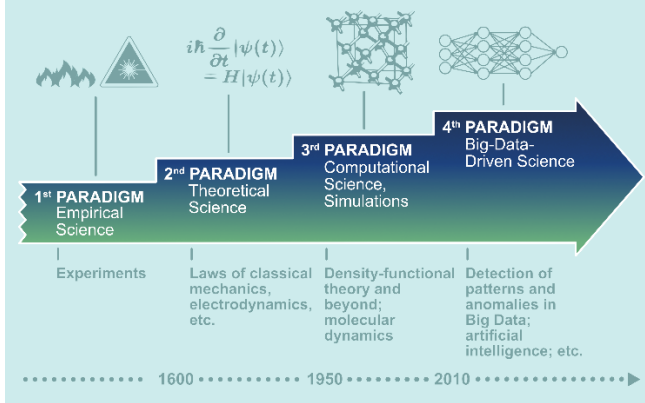

**Figure 1.** Development of the paradigms (new modes of thought) of materials science and engineering.

Nowadays, Big-Data and artificial intelligence revolutionize many areas of our life and the sciences. Materials science is not an exception (for a few examples, see Refs. 18, 19, 20, 21, 22, 23, 24, 25, 26, 27, 28). In fact, data-driven science is becoming the forth paradigm of materials research as sketched in Fig. 1.[29] While experimental investigations have been carried out since the stone and copper age, scientists of the 16[th] century started to describe physical relations by equations. Thus analytical equations became a central instrument of theoretical physics (symbolized by the Schrödinger equation in Fig. 1.). Obviously, with the event of the second paradigm, the first one, die empirical and experimental sciences did not become obsolete but they were complemented by an important, novel methodology. The 1950s marked the beginning of computational materials science and simulations, the third paradigm. Within this framework, computer experiments and simulations became possible, with the corresponding results being analyzed und interpreted like measured ones.

The vision of NOMAD was and is to establish the fourth paradigm[29] in computational materials science (cf. Fig. 1), i.e., the recognition that many properties of materials cannot be described by a closed mathematical form as they are determined by several multi-level, intricate theoretical concepts. Big-Data contain correlations, reflected in terms of structure and patterns in the data that are not visible in small data sets. Machine learning[30], compressed sensing[27,28],

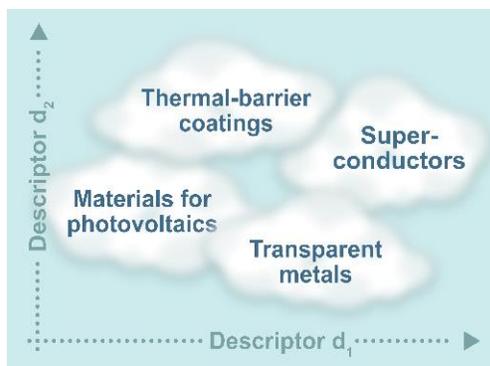

**Figure 2.** Big-Data contain correlations and structure that are not visible in small data sets. Finding descriptors that determine a specific property or function of a materials is a crucial challenge. Once being in place, we will be able to machine learn the data and eventually draw maps of subgroup discovery[31] and other methods of Big-Data Analytics and *Data Mining* can identify these patterns and transform them into useful information (cf. Fig. 2).

Thus, a key aspect of NOMAD's vision is to draw "materials maps", that tell us where in the "structural and chemical compound space" we can find materials that are useful for a particular application, as sketched in Figure 2. Works like those of Phillips and Van Vechten[32,33] and Ashby plots[34] are role models though these "materials maps" employ "(*x, y*) coordinates" that need knowledge about the described material. Not for materials, but for atoms, the Periodic Table of Elements is a fascinating early role model for our vison of such maps, where totally unknown elements were identified in terms of white spots in the table, and even their rough chemical properties could be predicted. Amazingly, this table was developed before any understanding of quantum mechanics and the shell structure of the electrons. Materials maps will, for sure, be complex and multidimensional. The crucial challenge in this context is finding the actuators behind the materials properties and functions, i.e., the proper descriptors. These need to be based on atomic properties and "collective" quantities that are much less involved to determine than those to be predicted. Turning this vison into reality requires novel approaches as mentioned above, and we need an extensive data management: Similarly to the establishment of a stable internet, we now need to install a comprehensive Big-Data infrastructure.

Finally, we stress that the success of novel data-mining tools strongly depends on the data quality[35,36]. The NOMAD Laboratory[10] has set out to address all these aspects of computational and data-driven materials science, providing a true platform for *Open Materials Science*. It starts with the *NOMAD Repository[37]*, by now the world-wide largest raw-data collection of its kind; in the *NOMAD*

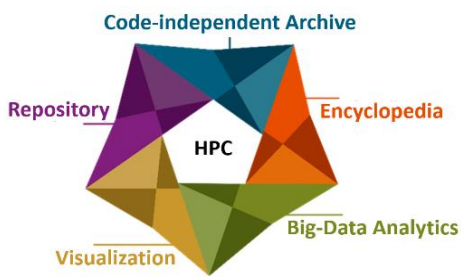

**Figure 3.** Structure of the NOMAD CoE[10]. More information can be found at a three-minute video https://youtu.be/yawM2ThVlGw.

*Archive[16]* the raw data are brought to a normalized form; the *NOMAD Encyclopedia[38]* is the "face" of the project, displaying the content of the vast amount of data, thereby also using *Advanced Visualization* tools; while the *NOMAD Analytics Toolkit* demonstrates novel data-mining methodology for data-driven research. The overall concept of the NOMAD CoE, as sketched in Figure 3 and is described in more detail below.

**The NOMAD Laboratory**

The Novel Materials Discovery Laboratory – in short NOMAD CoE[10] – is a European Centre of Excellence (CoE) which was established in fall 2015. Eight complementary research groups along with four high-performance computing (HPC) centers form the synergetic core of this CoE. This group also reflects that the CoE is part of the Psi-k[39], CECAM[40] , and ETSF[41] communities.

In short, the NOMAD CoE creates, collects, processes, stores, cleanses, and visualizes computational materials science data, computed by the most important materials-science codes available today. Most important, the NOMAD CoE develops innovative tools for mining this data in order to find structure, correlations, and novel information that could not be discovered from studying smaller data sets. The big picture is to advance materials science by enabling researchers in basic science and engineering to understand materials data, identify new materials and physical phenomena, and thus help industry to improve existing and develop novel products and technologies.

***The NOMAD Repository***

The first level of the NOMAD CoE is the NOMAD Repository. It contains the input and output files from several million high-quality calculations by now, and the data volume made available through the NOMAD Repository is rapidly increasing. In fact, the computational materials-science community uses millions of CPU hours every day in HPC centers worldwide. The NOMAD data collection comprises calculations that have been produced with any of the leading electronic-structure codes and increasingly also with codes from quantum chemistry. Presently, NOMAD supports about 40 codes (see also NOMAD Archive, below), and less-frequently used codes will be added on demand. Thus, the NOMAD concept and praxis is orthogonal to other data collections, because the NOMAD Repository is not restricted to selected computer codes or closed research teams but serves the entire community with its ecosystem of very different computer codes.

The NOMAD Repository contains not only plenty of calculations performed by individual researchers from all over the world but also the data of the most important computational materials databases worldwide: Examples are AFLOW[42], OQMD[43], and The Materials Project[44]. As such NOMAD copes with the increasing demand of storing scientific data and making them available for longer periods, as required by many funding agencies worldwide. NOMAD keeps scientific data for at least 10 years for free. That way, NOMAD also facilitates research groups to organize their data for their own groups and to share and exchange their results between two or more groups, and to recall what was actually done some years ago. The NOMAD Repository is the only repository in materials science so far to be recommended by Scientific Data[45].

Results are accepted in their raw format as produced by the underlying code. The only condition is that the list of authors is provided, and code and code version can be retrieved from the uploaded files, and the input and output files have to be complete. Data can be restricted to the owner or made available to other people (selected by the owner). After a maximum period of three years though, all data become open access. For downloading data, not even registration is required. As noted above, the NOMAD concepts are 100% FAIR[11,12,14]. The NOMAD Repository is online and open since beginning of 2014 and has since then enabled a cultural shift towards open data and thus *Open Science*[46] in the computational materials research. For getting a quick introduction into the NOMAD Repository, we recommend to watch a short movie[47].

### *The NOMAD Archive*

As the NOMAD Repository data is generated by many different computer codes, it is very heterogeneous. We have developed ways to convert the existing open-access data of the NOMAD Repository into a common, code-independent format, developing numerous parsers and creating the NOMAD Archive. The NOMAD Archive consists of the open-access Repository data, converted to a normalized form. This ensures that data from different sources can be compared and, hence, collectively operated upon by various NOMAD tools. A clear and usable metadata definition is a prerequisite for this normalization step to a code-independent format The development of the NOMAD Meta Info[16] was, indeed, a challenge. When NOMAD started, practically no metadata existed for describing materials data but they are of crucial importance also in view of the fact that physics, chemistry, and materials science may use one term for different quantities or different terms for one and the same property.

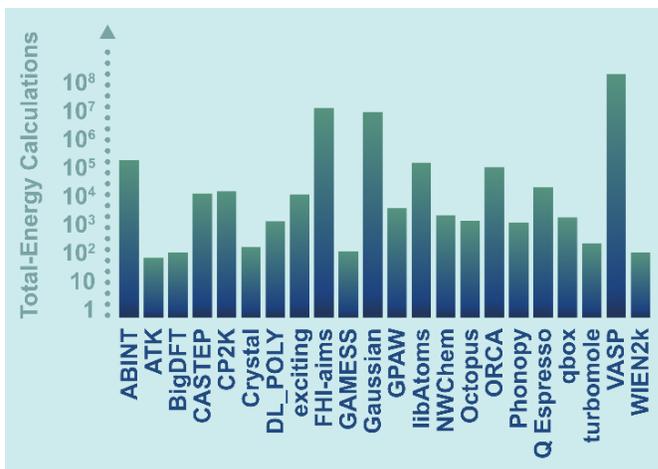

**Figure 4.** The NOMAD Laboratory supports all important codes in computational materials science. The figure shows the number of uploaded open-access total-energy calculations at the NOMAD Repository [1] as of March 15, 2018. The abscissa shows the various codes with more than 80 uploads. The total number of open-access total-energy calculations at the NOMAD Repository is more than 50 million, corresponding to billions of CPU-core hours.

Figure 4 shows, how many total-energy calculations produced by different codes were found in the NOMAD Repository and Archive by March 15, 2018.

### *The NOMAD Encyclopedia*

Knowledge and understanding of materials is based on their characterization by a variety of measured and/or computed properties. This includes structural features, mechanical and thermal behavior, electronic and magnetic properties, the response to light, and more. For the computational side, the NOMAD CoE has created a data infrastructure for not only collecting and sharing data but let us see what all this data contain. Indeed, the variety of data uploaded to the NOMAD Repository contain a lot of information that has not even been explored by the people creating the data. The NOMAD Encyclopedia is a web-based infrastructure that makes the many millions of calculations accessible. Its graphical user interface (GUI) provides a materials-oriented view on the computational materials data of the NOMAD Archive, displaying all properties of a given material which have been computed all over the world. In other words, it represents a user-friendly, public access point to the extensive knowledge contained in the NOMAD Archive. It not only helps us to search for the properties of a large variety of materials, but also directly see the spread of results and, as for instance, the impact of a density functional on a given feature. So far, we process structural and electronic properties and thermal behaviors, and alike for bulk materials and low-dimensional systems. Very shortly, the Encyclopedia will handle molecules, surfaces and adsorbate systems, the response to external excitations, elastic properties, Fermi surfaces, molecular-dynamics, and more.

Furthermore, the Encyclopedia provides a material classification system, links to external sources, and last but not least also an error-reporting tool for the case of problematic data. Should there be a dataset or a graph that does not appear to be correct, the user can, with the help of a simple menu, let us know about it.

### The NOMAD Visualization Tools

Seeing helps understanding. Consequently, NOMAD has developed an infrastructure for remote visualization of the multi-dimensional NOMAD data. We provide a centralized service that enables users to interactively perform comprehensive data visualization tasks on their computers without the need for specialized hardware or software installations. It allows researchers to conveniently perform graphical analyses of complex and multi-dimensional, time-dependent data from electronic-structure simulations, together with molecular structures. A special focus is laid on virtual-reality (VR) for interactive data exploration. Users have access to data and tools using standard devices (laptops, smartphones), independent of their location. Such VR enhances training and dissemination and even were a great success when presented to the general public. As an example, we note that 360-degree movies can be even watched with simple Google cardboard glasses as demonstrated, e.g. for $CO_2$ adsorption on $CaO$[48] and excitons in $LiF$[49]. The latter, being six-dimensional objects, cannot easily be visualized in an insightful manner otherwise. Taking the position of an electron or a hole, VR allows for inspecting the space of its counterpart.

### *The NOMAD Analytics Toolkit*

Having the data of the NOMAD Repository and Archive at hand, we now address the question how this data can be turned into knowledge and understanding. Let us emphasize first that the amount of available data is huge (billions of results) but compared to the immensity of possible materials and the very many intricate processes that determine the materials properties, the coverage of the structural and chemical compound space is still shallow.

Interestingly, this immensity of possible materials is sparsely populated when the focus is on selected properties or functions, e.g. systems with a certain band gap and effective mass that are stable under ambient conditions. Thus, we are looking for some needles in a haystack, and if our tools are too approximate we may describe the hay but may miss the needles. Our aim is to develop Big-Data analytics tools that will help to sort all of the available materials data to identify trends and anomalies and to build maps that also cover presently not yet synthesized materials, as sketched in Fig. 2.

Machine learning approaches are non-linear fits of a large pool of data. They will work, when there is enough data. However, for most materials properties this is often not the case, and the question has been raised if there will be ever enough data. The approach will work on less data if based on a clever descriptor (a set of descriptive parameters) that ensures that the data is arranged in a somewhat smooth manner.[27,28] Obviously, finding the descriptor may require domain knowledge, particularity when a highly accurate and predictive description is required. Progress in this direction has been significant, in particular exploiting the mentioned sparsity by compressed sensing. In general, the systematic search for descriptors for materials science is still in its infancy.

In this spirit, the NOMAD CoE is developing an "Analytics Toolkit". The overarching topics that are currently addressed are (i) crystal-structure prediction; (ii) scanning for good thermoelectric materials; (iii) finding better materials for heterogeneous catalysis, (iv) searching for better materials for optoelectronics and photovoltaics, (v) analyzing alloys and their plasticity, (vi) predicting topological insulators, and more. A number of prototype applications of our data-mining approaches are already publicly available.[50]

Let us demonstrate the mentioned concept of finding descriptors and building "maps of materials" by a recent example that employed compressed sensing for the descriptor identification. Compressed sensing originates from signal processing and finds a low dimensional representation for a complex signal. In materials science, it has been used to identify the key physical actuators that are behind materials properties and it identifies a few leading descriptor equations out of a huge number of candidates.[27,28] Specifically, we here summarize a recent work by Acosta and coworkers[51] who used the approach to build a materials map for two-dimensional honeycomb structures in order to analyze and identify two-dimensional topological insulators (also called quantum spin Hall insulators, QSHIs). A number of 220 functionalized honeycomb-lattices that are isoelectronic to functionalized graphene were calculated. These materials are built of group IV, or III-V, or IV-VI, elements, and all atoms are bonded to a group VII element. Besides confirming the QSHI character of well-known materials, the study revealed several other yet unreported QSHIs. Using a recently introduced method [28], called SISSO (sure independence screening and sparsifying operator), the authors then offered 10 million candidate descriptors to identify the best low-dimensional descriptor. Figure 6 shows the corresponding "materials map" defined by the two-dimensional descriptor. Analysis of these descriptors (not shown here) yields fundamental insights in the mechanisms driving topological transitions. Furthermore, the map predicts several new QSHIs that were not part of the calculated materials.[51]

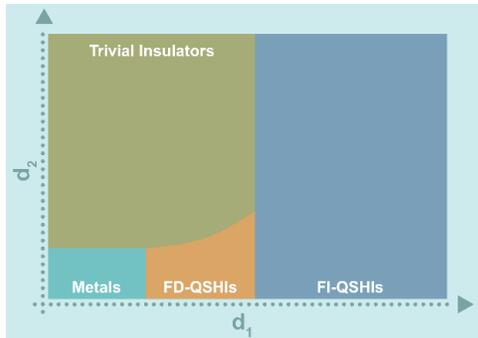

**Figure 5.** A map of materials classifying two-dimensional functionalized honey-comb structures as functionalization-independent Quantum Spin Hall Insulators (QSHIs), functionalization-dependent QSHIs, metals, and trivial insulators.

**Summary and outlook**

The growth of data from simulations (and experiments) is expanding beyond a level that is addressable by established scientific methods. The so-called "4 V challenge" of Big-Data –**V**olume (the amount of data), **V**ariety (the heterogeneity of form and meaning of data), **V**elocity (the rate at which data may change or new data arrive), and **V**eracity (uncertainty of the data quality) – is clearly becoming eminent also in materials science. Controlling this massive amounts of data, on the other hand, sets the stage for explorations and discoveries. Novel data-mining technology can find patterns and correlations in data that can't be seen in small data sets or standard tools. As such, data-driven materials research is adding a new research paradigm to our scientific landscape. All this is in the heart of the NOMAD CoE.

Handling experimental data in a similar fashion like we treat computational data – the logical next step - will make the data challenge even more intricate. Experimental apparatus nowadays produce terabytes of data per day in every mid-sized facility. Taking transmission-electron microscopes as an example, not only can detectors acquire thousands of datasets per second with millions of data points at once, also multiple detectors can operate in parallel for multi-signal acquisition. This leads to data creation at typical rates of 30 GB per second – orders of magnitude faster than existing data processing and analysis tools can cope with. If this mass of data is to be effectively captured, managed, navigated and exploited, novel HPC solutions for high-performance data processing, data storage, data analytics, information retrieval, and user experience are essential. Notably, experimental data stem from various measurement techniques and instruments with different resolutions and other experimental parameters. Another crucial issue is the quality of experimental samples, which is often not sufficiently known or characterized. In fact, the sample's quality depends very much on its history. Metadata about the experiment, the sample and the data must be defined if the experimental data is re-used or combined with other data. This is hardly done to date; The NOMAD team together with experimental colleagues[52] is doing first steps in this directions.

**Acknowledgments**

This work has received funding from the European Union's Horizon 2020 research and innovation programme, grant agreement No 676580, the NOMAD Laboratory CoE and No 740233, TEC1P. Support from the Einstein Foundation Berlin is appreciated. We thank Christoph Koch for in-depth discussions and information about transmission electron microscopy, Peter Wittenburg for clarification of the FAIR concept, and the whole NOMAD team for the invaluable effort to build the entire NOMAD infrastructure and its services.

**Author biographies**

Claudia Draxl
Physics Department and IRIS Adlershof
Humboldt-Universität zu Berlin, Berlin, Germany
+49 30 2093 66363 Phone
+49 30 2093 66361 Fax
Email: claudia.draxl@physik.hu-berlin.de
Claudia Draxl is full professor at the Humboldt-Universität and Max-Planck Fellow at the Fritz-Haber-Institut Berlin, Germany. She studied physics and mathematics at the University of Graz, where she got her PhD in solid-state theory. Among other prizes, she won the Ludwig Boltzmann Award of the Austrian Physical Society and was conferred a honorary doctorate of Uppsala University (Sweden).

Her research is dedicated to the understanding of conventional and organic semi-conductors and interfaces thereof, complex alloys, 2D systems and more. A major focus is excited states, using and developing techniques beyond density functional theory. Another focus point concerns data-driven research. She is author of about 250 publications in peer-reviewed journals.

Matthias Scheffler
Fritz-Haber-Institut der Max-Planck-Gesellschaft
+49 30 8413 4711 Phone
+49 30 8413 4701 Fax
Email: scheffler@fhi-berlin.mpg.de

Matthias Scheffler is director at the Fritz Haber Institute of the Max Planck Society. He is also honorary professor of theoretical condensed matter physics at all three Berlin universities and distinguished visiting professor for materials science and engineering at the UC Santa Barbara and Hokkaido University (Sapporo). His research is concerned with basic aspects of physical and chemical properties and processes at bulk materials, surfaces, interface, defects, clusters, and nanostructures. He developed various methods for first-principles computations of the electronic structure and systematic hybrid approaches of electronic-structure theory, thermodynamics, and statistical mechanics. Since recent years the emphasis of his studies is on big-data-driven materials science.

---

accepted language for data representation; 2) The vocabulary of metadata and data follow FAIR principles; 3) Metadata and data include qualified references to other (meta)data and to the authors who created the results.